# Image potential states as quantum probe of graphene interfaces

Sangita Bose[1], Vyacheslav M. Silkin[2,3,4], Robin Ohmann[1], Ivan Brihuega[1], Lucia Vitali[1], Christian H. Michaelis[1], Pierre Mallet[5], Jean Yves Veuillen[5], M. Alexander Schneider[6], Evgueni V. Chulkov[2,3,7], Pedro M. Echenique[2,3,7] and Klaus Kern[1,8]

[1]Max Planck Institut für Festkörperforschung, Heisenbergstrasse 1, D-70569 Stuttgart, Germany

[2]Donostia International Physics Center (DIPC), Paseo de Manuel Lardizabal 4, E-20018 San Sebastian, Basque Country, Spain

[3]Dpto. de Física de Materiales, Facultad de Ciencias Químicas, Universidad del País Vasco (UPV/EHU), E-20080 San Sebastian, Basque Country Spain

[4]IKERBASQUE, Basque Foundation for Science, 48011, Bilbao, Spain

[5]Institut Néel, CNRS et Université Joseph Fourier, BP 166, F-38042 Grenoble, Cedex 9, France

[6]Lehrstuhl f. Festkörperphysik, Universität Erlangen-Nürnberg, Staudtstrasse 7, D-91058 Erlangen, Germany

[7]Centro de Física de Materiales CFM (Centro Mixto CSIC-UPV/EHU), Edificio Korta, Avenida de Tolosa 72, E-20018 San Sebastian, Basque Country Spain

[8]Institut de Physique de la Matière Condensée, Ecole Polytechnique Fédérale de Lausanne, CH-1015 Lausanne, Switzerland

E-mail : sangita.bose@fkf.mpg.de

**Abstract:** Image potential states (IPSs) are electronic states localized in front of a surface in a potential well formed by the surface projected bulk band gap on one side and the image potential barrier on the other. In the limit of a two-dimensional solid a **double** Rydberg series of IPSs has been predicted which is in contrast to a single series present in three-dimensional solids. Here, we confirm this prediction experimentally for mono- and bilayer graphene. The IPSs of epitaxial graphene on SiC are measured by scanning tunnelling spectroscopy and the results are compared to *ab-initio* band structure calculations. Despite the presence of the substrate, both calculations and experimental measurements show that the first pair of the double series of IPSs survives, and eventually evolves into a single series for graphite. Thus, IPSs provide an elegant quantum probe of the interfacial coupling in graphene systems.

**1. Introduction**

Free standing graphene (FSG) exhibits a large number of novel properties arising primarily from its two-dimensionality [1,2,3], many of which are preserved in epitaxial graphene [4,5,6]. A remarkable consequence of the two-dimensional (2D) character, is the occurrence of *two series of image-potential states* (IPSs) at the Brillouin zone center as predicted recently by some of us from *ab initio* band structure calculations in FSG [7] which is in contrast to surfaces of three-dimensional (3D) solids where only one series of such states exists [8,9,10]. The wave functions of these states for the two series have opposite parity with respect to the reflection at the graphene plane. The state with a symmetric wave function, $n^+$, is lower in energy compared to the state with anti-symmetric wave function, $n^-$, for each order, $n$. Due to its large spatial expansion into the vacuum, these states provide a very sensitive tool for the determination of the environment around the carbon sheet. On bilayers (BL) of FSG the same calculations show the *two series of IPSs* but the energy difference between them is smaller than between the corresponding states of the dual series in the monolayer (ML). Therefore, it is expected that the *two series of IPSs* in FSG merge into a single series with increasing number of carbon layers when one goes from 2D graphene to 3D graphite.

Epitaxial graphene grown on SiC with a varying number of carbon layers provides a unique system to measure the evolution of IPSs with increasing layer thickness [4,6]. In addition it opens the possibility to explore the effect of an underlying substrate on the two series of IPSs in FSG. The measurement of IPSs on epitaxial graphene provides a way to determine the electronic coupling between the graphene sheet and the SiC substrate underneath. Scanning tunneling spectroscopy (STS) in the resonant tunneling mode is a useful tool to probe electronic states extending into vacuum close to the Brillouin zone center [11,12]. In this method, electrons emitted from the tip interact with surfaces and interfaces at high voltages (> 2 V) giving rise to resonances which correspond to the Stark shifted IPSs.

In this paper we show through a measurement of Stark shifted IPSs by STS and band structure calculations using density functional theory on epitaxial graphene that the lowest symmetric and anti-symmetric states ($1^+$ and $1^-$), of the two Rydberg series of IPSs ($n^+$ and $n^-$) predicted for FSG indeed **persists** when graphene is grown on SiC, thereby indicating only a partial coupling to the underlying substrate. The states with higher quantum numbers (n >1) convert into the IPSs of the entire system (graphene + substrate), very similar to the formation of 'tubular IPSs' in carbon nanotubes [13] and 'super atom' molecular orbitals in fullerenes [14]. Interestingly, with increasing number of carbon layers, the two series evolve into a single series of IPSs for graphite. In addition the lowest symmetric state of the two series in graphene evolves into the interlayer band in graphite.

## 2. Methods

*2.1 Experimental methods*

The image potential states (IPSs) have been measured using a home-built scanning tunneling microscope operating at 1.4 K under ultra high vacuum conditions (UHV). The HOPG crystal used was cleaved *in situ* under (UHV) conditions (1 x $10^{-10}$ mbar). The epitaxial graphene used for the experiments was grown by thermal desorption of silicon on a 6H-SiC(0001) substrate [6, 15]. The samples were imaged at 4.2 K in constant current mode in the STM. The ML and BL terraces were identified using the different contribution of the interface states in the two, following the method used previously [15]. Scanning tunneling spectroscopic (STS) measurements of the IPSs were performed with constant current in a closed feedback loop where the distance-voltage (z(V)) characteristics and the dI/dV signal were simultaneously recorded [9,10,12]. All spectroscopic measurements have been measured with the modulation technique using a lock-in amplifier, where the voltage modulation to the sample bias voltage was 20 mV at a frequency of 1 kHz. The lock-in signal (dI/dV) shows peaks corresponding to the Stark shifted IPSs.

*2.2 Theoretical calculations*

Self-consistent density functional calculations of the electronic structure of the isolated ML and BL graphene sheets were performed using the plane-wave basis set with a cutoff of 50 Ry and a 36x36 $\mathbf{k}_{\parallel}$ grid for the density within a local-density approximation (LDA) with exchange-correlation potential of Ref. 16. Note that the two lowest-energy unoccupied states resembling the $1^+$ and $1^-$ IPSs appear in the LDA calculations below the vacuum level due to the unusually deep potential well in the vicinity of carbon-sheets [7]. As here we are primarily interested in the energy positions of two lowest IPSs in the vicinity of carbon sheets in the presence of a substrate and applied external electric field, the use of LDA potentials works reasonably well. The electron-ion interaction was described by a norm-conserving pseudopotential [17]. An external electric field was incorporated into the calculations via removing the corresponding amount of electric charge from the ML or BL carbon sheets into the vacuum.

**3. Results and discussions**

Figure 1(b) shows the Stark shifted IPSs measured on ML and BL epitaxial graphene as well as graphite using STS at a tunnelling current of 0.2 nA (Schematic diagrams and STM topographic images with the atomic contrast on the three surfaces are shown in Figure 1(a)). For epitaxial graphene, the IPSs were measured on the ML and BL (Figure 1(a)) under identical conditions, with the same 'microscopic tip'. For graphite, the measurement was carried out several times with the same current of 0.2 nA and the same initial set point voltage of 2.0 V. There are two distinct features in the measured IPSs on the three surfaces. (1) A strong peak appears at ~4 V and shifts to higher energies with increasing number of carbon layers. (2) A distinct 'hump' is observed in ML graphene at a bias voltage of ~3.3 V which is also faintly visible at the same energy in the case of BL graphene. We would like to point out that resonant states for bias voltages > 5 V extend deeper into the vacuum with a higher probability of the electron density close to the tip [11,18]. Since they are strongly influenced by the electric field of the STM tip, the spectroscopic information obtained from these states should be treated with caution [10,12]. Further we note that the two states appear at a much lower energy than expected for quantum-well states (QWS) in few layer graphene systems [19] and show different energy *vs* thickness dependence than expected for QWS. However among the higher lying resonances there might be contributions from the QWS, therefore we will focus on states with energetic position < 5 eV.

To understand the origin of the states in ML and BL epitaxial graphene, self-consistent calculations of the band structure were performed. The calculations have been performed in several steps. In the first step, local density approximation (LDA) calculations including the image-potential tail were done for FSG in the absence of an external electric field (the details of the calculations are given in Ref. [7]. These predicted the occurrence of the *two series of IPSs* in FSG. It is worth noting that Posternak *et al* had also shown from their LDA calculations on ML and BL graphene the presence of two unoccupied states at the Brillouin zone center [20]. They further showed that the bonding combination of two symmetric states of each ML gave rise to the symmetric state in BL while their anti-bonding combination gave rise to the anti-symmetric state. Consequently the anti-symmetric state with increasing number of carbon layers was claimed to evolve into the surface state of graphite [21] but later it was interpreted experimentally from multiphoton photoemission spectroscopy to evolve into the lowest IPS [22]. However, the complete *two series of IPSs* prevalent at each order could be obtained only from a band structure calculation using the image potential tail along with LDA in the vicinity of the graphene sheet.

As a next step, the influence of an external electric field ($E_{ext}$) on the IPSs of both ML and BL FSG was calculated. The external electric field was applied on one side in order to simulate the STM experiments where the tip voltage appears from one side (Potential diagram in Figure 2). The corresponding local parts of self-consistent potentials averaged in plane for $E_{ext}$ = 0, 0.1, 0.2, and 0.3 V/Å are shown in Figure 2. Finally, the effect of the substrate was included by adding a constant potential $V_s$ at a distance of $-z_s$ from the first carbon layer. $V_s$ is determined by the position of the upper edge of the substrate (SiC) potential ($E_{upper}$) with respect to the Fermi level. The position of the Fermi level was ascertained in ML and BL by the position of the Dirac point for different $E_{ext}$. The shift of the Fermi level for different $E_{ext}$ was determined from the evaluation of the corresponding doping level.

Interestingly, the calculations on graphene simulating the STM tip and the underlying substrate (as described above) show that the wave functions of only the first two IPSs (between 2.5-5 eV) in both ML and BL have opposite symmetry indicating the presence of only the lowest-energy ($1^+$ and $1^-$) states of the dual series. Consequently, the lower in energy of the paired states is symmetric ($1^+$ state) and the higher is anti-symmetric ($1^-$ state) (shown in Figures. 3(a) and 3(b)) similar to FSG (for comparison the partial densities in FSG of the $1^+$ and $1^-$ states are shown in Figures 3(a) and 3(b) by dashed lines). However for graphene (including a substrate potential), in contrast to FSG, the higher *n* levels (n > 1, $E_n$ > 5.0 eV) do not show any splitting and appear as resonant states (as seen from the calculations). The reason for this is that for higher *n* states, there is no space between the graphene and the substrate for the wave functions to be localized and therefore only the conventional image state series occurs with the main charge density being localized above the graphene sheet in the vacuum side. The resonance character of these states is explained by closure of the energy gap of the substrate at the corresponding energies [23,24]. If we plot the difference in the densities of the respective split states in ML and BL in the vacuum side (Figure 3(c)) for $E_{ext}$ = 0.1 V/Å, we observe that both the $1^+$ and $1^-$ states are expanded in vacuum more in case of the ML compared to the BL. This implies that the electric field will perturb the ML more and will result in a higher Stark shift of the IPSs compared to the BL.

To check whether the predicted two series of IPSs are observed in our experimental spectra of the Stark shifted IPS measured on ML and BL epitaxial graphene, we fitted the observed peaks between 2.5-5 eV (after subtracting the background) with a sum of two Gaussian functions. We obtained a very good fit in case of the ML (Figure 4(a)), ascertaining the presence of two distinct peaks in the energy range (2.5-5 eV) where the lowest states of the two series ($1^+$ and $1^-$) are expected. We attribute the 'hump' obtained at ~ 3.3 V of the ML to the $1^+$ state and the strong peak at ~4 V to the $1^-$ state. To investigate it further, we measured the IPSs in the ML with 'microscopically' different tips. For all tips we observe a hump and a strong peak at almost the same energy while the higher energy (> 5 eV) peak positions vary for different tips (Figure 5(a)). This suggests that the first two peaks can be attributed to the two series of IPSs. We further carried out a spatial mapping of IPSs on ML where the measurement was done across a line as shown in the topographic image of the ML terrace (Top of Figure 5(b)). The conductance image shown in the bottom panel of Figure 5(b) shows the dI/dV signal recorded as a function of bias voltage at different spatial points along the straight line on the ML terrace. From spatially resolved IPS-spectroscopy (Figure 5(b)) we find, that the intensities of the $1^-$ and $1^+$ state vary very little with atomic position (shown in Figure 5(b)) or varies with position relative to the Moiré-pattern (not shown, the pseudo 6X6 pattern obtained from the SiC buffer layer [15]) and hence show that the states are homogeneously extended over the full surface. We also obtain a good agreement between the energetic positions of these states in the ML as obtained from calculations using $E_{ext}$ = 0.1 V/Å and $z_s$ = -5 a.u. However, we obtain a huge difference in intensity in the two peaks (Figure 1(b)). The dominance of the $1^-$ peak in the dI/dV signal can be understood, if we look at the calculated partial densities of both the $1^+$ and $1^-$ states in the ML (Figure 3(c)). The $1^-$ state extends substantially deeper into the vacuum than the $1^+$ state. This will lead to a larger overlap with the tip wave function and enhance the tunnelling probability into this state resulting in the very large peak in the dI/dV signal observed at this bias voltage.

For the spectra measured on the BL, we observe two distinct differences from the ML. The first is that the hump at ~3.3 eV has become very weak (much poorer fits to the peak structure with the sum of two Gaussian functions, Figure 4(b)). From Figure 3(c) we see that the partial density of the $1^+$ state extends even less into the vacuum region in BL compared to the ML. Hence, the overlap with the tip wave function will be further reduced resulting in even lower intensity compared to the $1^+$ state in ML.

The second difference between ML and BL is that the $1^-$ peak is at a higher energy in BL compared to the ML. For similar external electric fields, we expect that the electric field will penetrate more in the ML compared to the BL since the substrate lies closer to the ML. This would imply that the states in the ML should be more Stark shifted than states in the BL, contrary to our observation. This leads us to the suggestion that the external electric field in BL is greater than 0.1 V/Å (value in the ML). This can indeed be seen from the measured height between the ML and BL at the same current and bias voltage, where the tip appears closer to the surface in the BL signifying a higher electric field compared to the ML (Figure 6).

From Figure 3(b), it also emerges that for a BL, some charge density gets trapped in between the carbon layers. Interestingly, this trapping of partial density in between the carbon layers should eventually evolve into an interlayer band with increasing number of carbon layers as observed in graphite which has its band minimum at 3.5 eV [19,22]. To see if there is a contribution to the n = 1 IPS in graphite from the interlayer band, we fit the peak between 3-5 eV with a sum of a Gaussian and a broad edge step function $y = \frac{A}{1 + \exp(-(V - V_{max})/B)}$, where A and B are fitting parameters and $V_{max}$ is the voltage where the peak has its maximum. We obtain a reasonable fit (Figure 4(c)), indicating that the n = 1 IPS has contributions from a second state located at the same energy (3-5 eV) which we attribute to the interlayer band.

In summary our measurements of IPSs by STM in epitaxial graphene provide a new way to determine the electronic coupling between graphene and an underlying substrate and also between the graphene layers. The observation of the lowest members of two series of IPSs demonstrates that epitaxial graphene is similar in character to pristine graphene despite the disturbance caused by the substrate and the strong electric field due to the STM tip. Additionally, we also observe that the lowest symmetric IPS ($1^+$) in graphene evolves into the interlayer band of graphite with increasing number of carbon layers, while the upper anti-symmetric IPS ($1^-$) evolves into the n = 1 image state of graphite.

**Acknowledgements**

S.B. would like to thank the Alexander von Humboldt foundation and I.B. the Marie Curie action program for support. The work was partially supported by the University of the Basque Country (Grant No. GIC07IT36607), Departamento de Educación del Gobierno Vasco, and the Spanish Ministerio de Educación (Grant No. FIS200766711C0101).

**Figure Captions:**

**Figure 1.** (a) Schematic of monolayer (ML), bilayer (BL) epitaxial graphene and graphite with the STM topographic images showing the atomic contrast in each system. The images were taken at I = 0.1 nA and sample bias of 2 mV. STM topographic image of epitaxial graphene on SiC showing a ML-BL terrace. Sample bias is +0.7 V. (b) dI/dV spectra obtained by lock-in technique for ML, BL and graphite. The tunneling current was 0.2 nA, the initial set point voltage = 2 V and the voltage modulation was 20 mV. The dashed lines are the Gaussian fits to the peaks. Clearly, in ML and BL there are two peaks in between 2.5-5 eV, while in graphite there is only 1 peak in the same energy range.

**Figure 2.** Potential diagram in presence of external electric field for epitaxial graphene including a substrate potential. Note that in the E = 0 V/Å case the LDA potential was modified for positive distances beyond 3 a.u. according to Ref. [7] in order to reproduce the image-potential tail.

**Figure 3.** Plot of the calculated partial density with distance for the first two IPSs (a)-(b) ML and BL epitaxial graphene are shown in solid lines. The vertical solid lines denote the position of the carbon plane. The dashed lines are for ML and BL FSG (The curves have been vertically displaced for clarity). Here 0 a.u. denotes the position of the top graphene layer. (c) Difference in the density in the vacuum region between ML and BL graphene (simulating the substrate) at an external electric field of E = 0.1 V/Å.

**Figure 4.** (a)-(b) Fit of the peak structure between 2.5-5 V with a sum of two Gaussian functions in ML and BL respectively showing clearly the split states (open circles are the raw data and the solid lines are the fits) (c) Fit of the raw data (open circles) for graphite between 3-5 V with a sum of a Gaussian and a broad-edge step function (solid line).

**Figure 5.** (a) Effect of the STM tip on the IPSs of epitaxial ML graphene. For clarity the curves are vertically displaced. (b) Spatial mapping of IPSs on ML graphene. The measurement was done over the line shown in the topographic image of the ML terrace (Top). Both the $1^+$ and $1^-$ states are visible in the spatial map of the IPSs (Bottom).

**Figure 6.** (a) Topographic STM image of the interface between ML and BL at a bias voltage = -0.6 V and I = 0.2 nA. **(b)** *z vs* Bias voltage in the z-V scans taken in the ML and BL with the same tip across the interface. The tip moves closer to the BL surface compared to the ML surface signifying a higher electric field. The zero in the y scale indicates the initial tip-sample distance determined by the initial tunneling conditions on the ML surface.

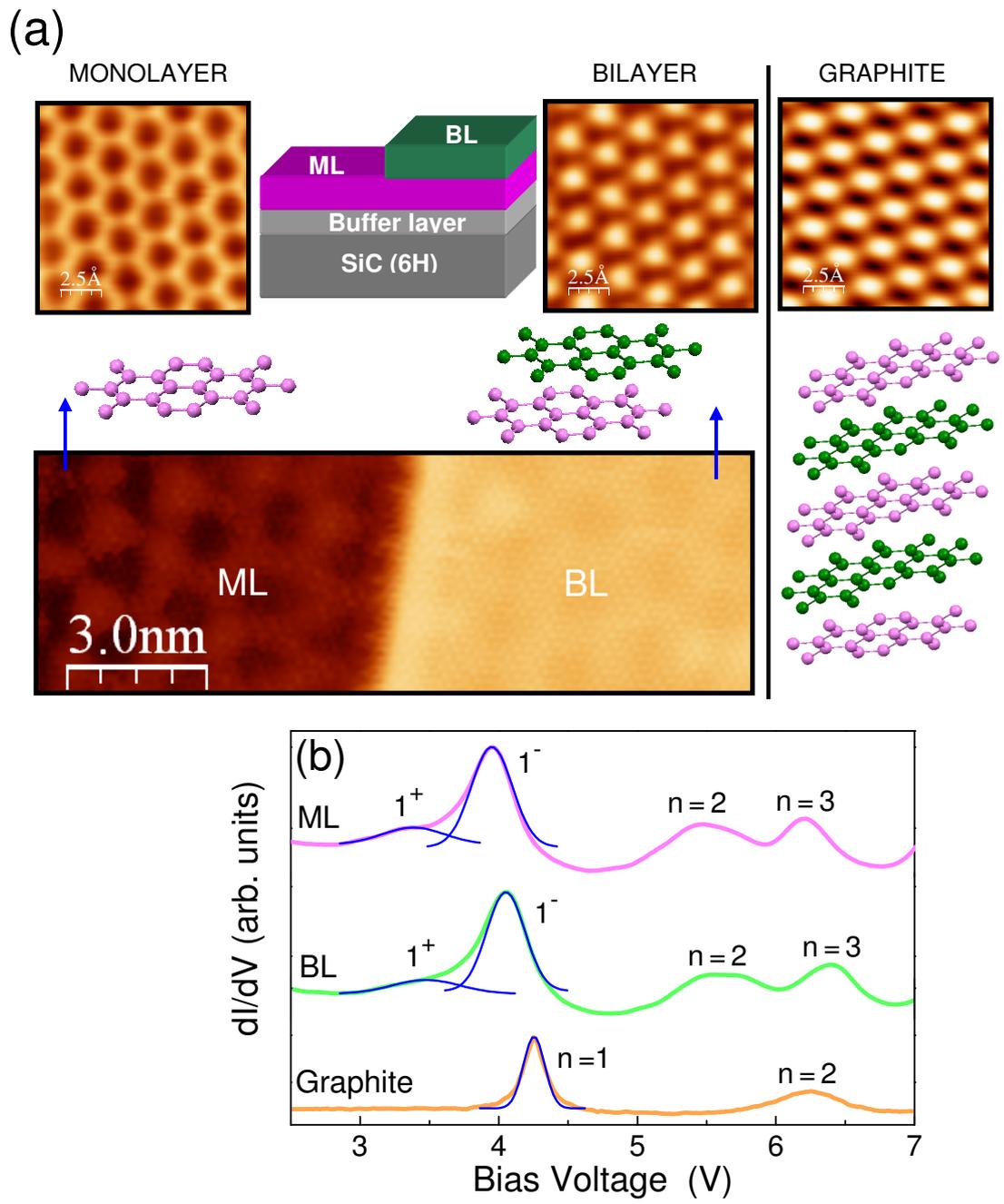

Figure 1

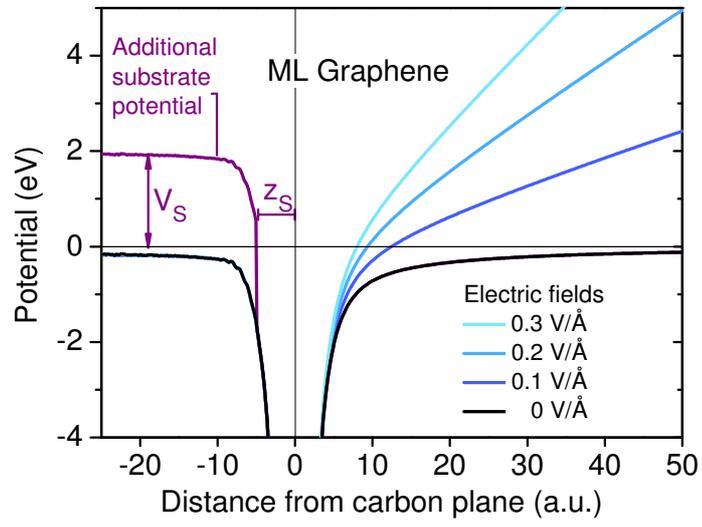

Figure 2

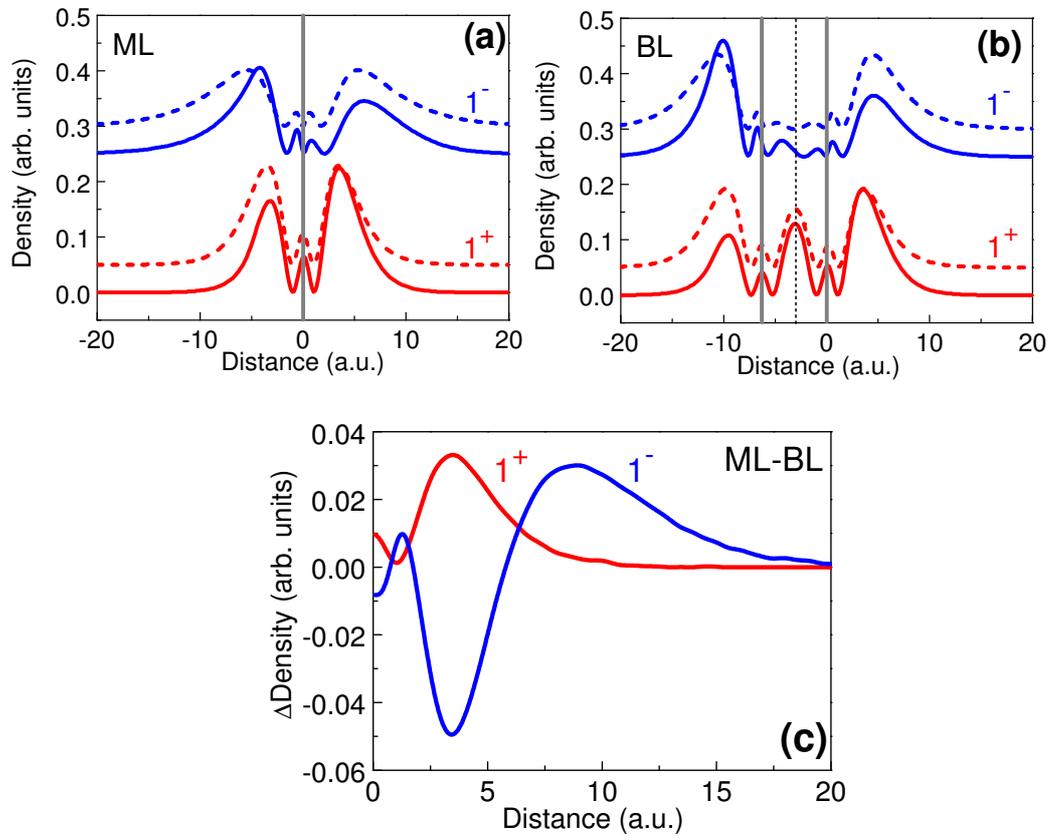

Figure 3

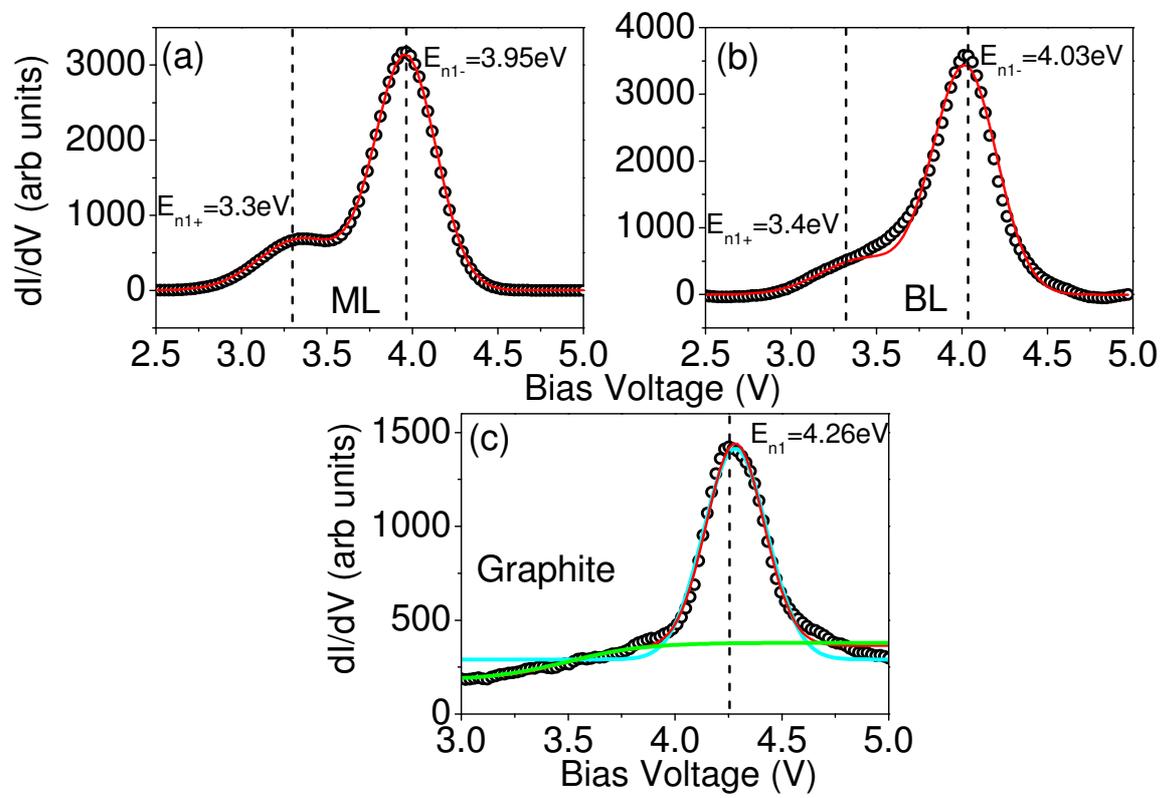

Figure 4

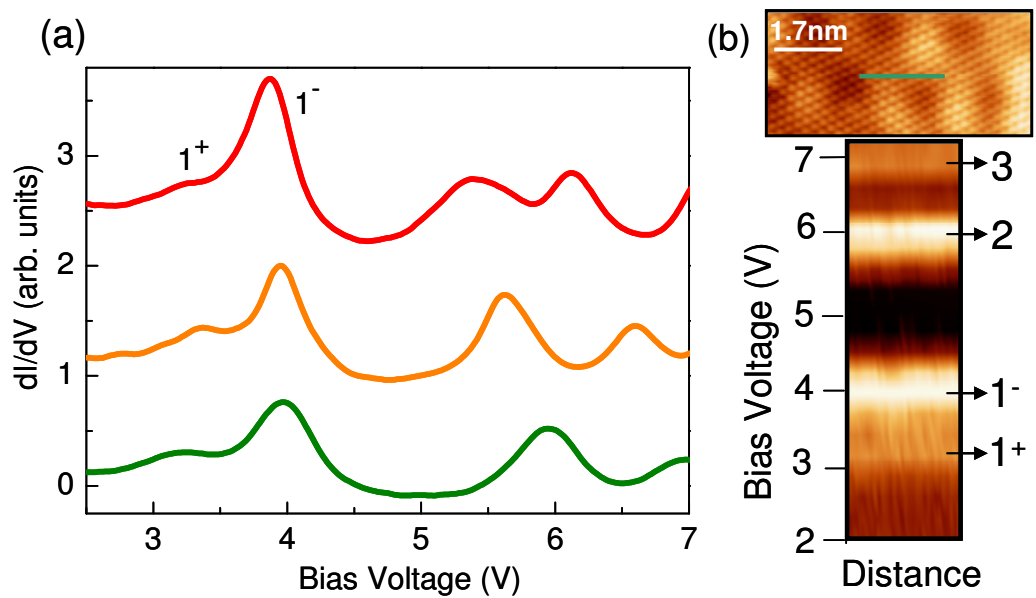

Figure 5

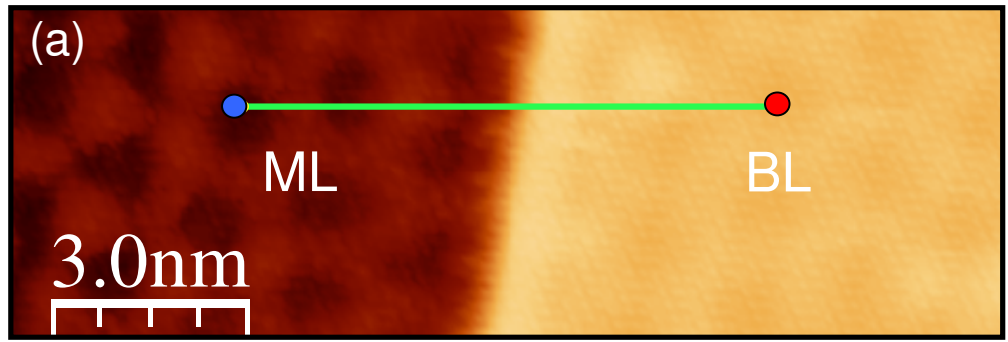

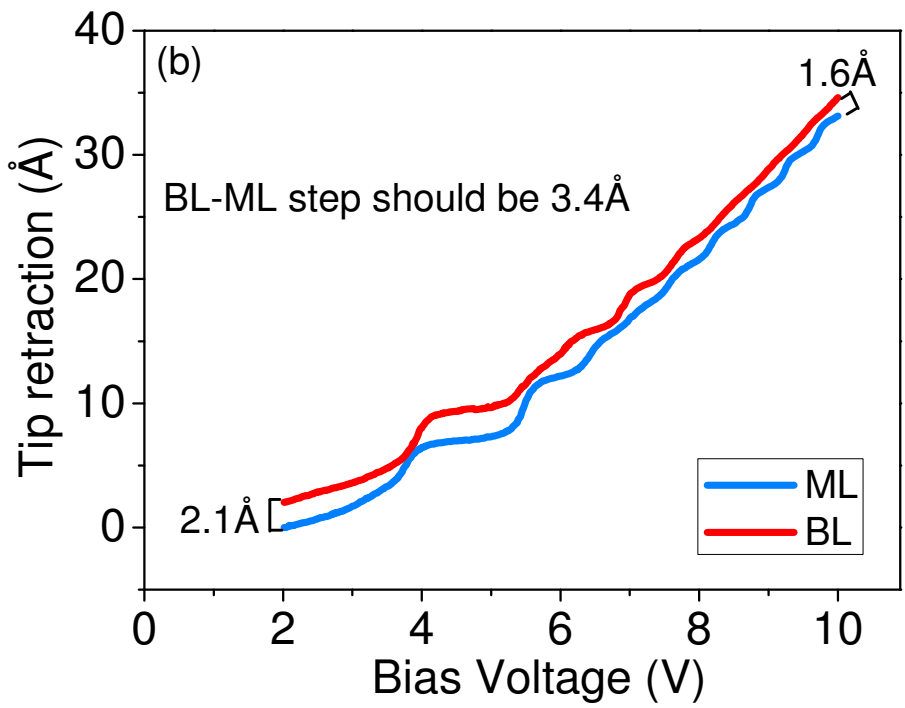

Figure 6

**References:**

[1] Novoselov K S, Geim A K, Morozov S V, Jiang D, Zhang Y, Dubonos S V, Grigorieva I V and Firsov A A 2004 *Science* **306** 666

[2] Morozov S V, Novoselov K S, Katnelson M I, Schedin F, Ponomarenko L A, Jiang D and Geim A K 2006 *Phys. Rev. Lett.* **97** 016801

[3] Zhang Y, Tan Y W, Stormer H L and Kim P 2005 *Nature* **438** 201

[4] de Heer W A, Berger C, Wu X, First P N, Conrad E H, Li X, Li T, Sprinkle M, Hass J, Sadowski M L, Potemski M and Martinez G 2007 *Solid State Communications* **143** 92

[5] Bostwick A, Ohta T, Seyller T, Horn K and Rotenberg E 2007 *Nature Phys.* **3** 36-40

[6] Brihuega I, Mallet P, Bena C, Bose S, Michaelis C, Vitali L, Varchon F, Magaud L, Kern K and Veuillen J Y 2008 *Phys. Rev. Lett.* **101** 206802

[7] Silkin V M, Zhao J, Guinea F, Chulkov E V, Echenique P M and Petek H 2009 *Phys. Rev. B* **80** 121408

[8] Hofer U, Shumay I L, Reuss C, Thomann U, Wallauer W and Fauster T 1997 *Science* **277** 1480

[9] Binnig G, Frank K H, Fuchs H, Garcia N, Reihl B, Rohrer H, Salvan F and Williams A R 1985 *Phys. Rev. Lett.* **55** 991

[10] Dougherty D B, Maksymovych P, Lee J, Feng M, Petek H and Yates J T 2007 *Phys. Rev. B* **76** 125428

[11] Ploigt H C, Brun C, Pivetta M, Patthey F and Schneider W D 2007 *Phys. Rev. B* **76** 195404

[12] Wahl P, Schneider M A, Diekhöner L, Vogelgesang R and Kern K 2003 *Phys. Rev. Lett.* **91** 106802

[13] Zamkov M, Woody N, Bing S, Chakraborty H S, Chang Z , Thumm U and Richard P 2004 *Phys. Rev. Lett.* **93** 156803

[14] Feng M, Zhao J and Petek H 2008 *Science* **320** 359

[15] Mallet P, Varchon F, Naud C, Magaud L, Berger C and Veuillen J Y 2007 *Phys. Rev. B* **76** 041403

[16] Ceperley D M and Alder B J 1980 *Phys. Rev. Lett.* **45** 566

[17] Troullier N and Martins J L 1991 *Phys. Rev. B* **43** 1993

[18] Pascual J I, Corriol C, Ceballos G, Aldazabal I, Rust H P, Horn K, Pitarke J M, Echenique P M and Arnau A 2007 *Phys. Rev. B* **75** 165326

[19]  Hibino H, Kageshima H, Maeda F, Nagase M, Kobayashi Y and Yamaguchi H 2008  *Phys. Rev. B* **77**, 075413

[20] Posternak M, Baldereschi A, Freeman A J, Wimmer E and Weinert M 1983 *Phys. Rev. Lett.* **50** 761

[21] Fauster T, Himpsel F J, Fischer J E and Plummer E W 1983 *Phys. Rev. Lett.* **51** 430

[22] Lehmann J, Merschdorf M, Thon A, Voll S and Pfeiffer W 1999 *Phys. Rev. B* **60** 17037

[23] Wenzien B, Käckell P, Bechstedt F and Cappellini G 1995 *Phys. Rev. B* **52** 10897

[24]  Varchon F, Feng R, Hass J, Li X, Nguyen B N, Naud C, Mallet P, Veuillen J Y Berger C, Conrad E H and Magaud L 2007 *Phys. Rev. Lett.*  **99** 126805